# Self-Assembled Monolayer Piezoelectrics: Electric-Field Driven Conformational Changes

Xinfeng Quan,[1] Jeffry D. Madura,[2] Geoffrey R. Hutchison[3,*]

[1] Sichuan University-Pittsburgh Institute, Chengdu, Sichuan, CN 610065

[2] Department of Chemistry & Biochemistry, Duquesne University, 600 Forbes Ave. Pittsburgh, Pennsylvania 15282.

[3] Department of Chemistry and Department of Chemical Engineering, University of Pittsburgh, 219 Parkman Avenue, Pittsburgh, Pennsylvania 15260.

*Correspondence to: geoffh@pitt.edu

**ABSTRACT** We demonstrate that an applied electric field causes piezoelectric distortion across single molecular monolayers of oligopeptides. We deposited self-assembled monolayers ~1.5 nm high onto smooth gold surfaces. These monolayers exhibit a strong electromechanical response that varies linearly with applied 1-3V bias (i.e., a converse piezoelectric effect), measured using piezoresponse force microscopy (PFM). The response is markedly greater than control experiments with rigid alkanethiols and correlates with surface spectroscopy and theoretical predictions of conformational change from applied electric fields. Unlike existing piezoelectric oxides, our peptide monolayers are intrinsically flexible, easily fabricated, aligned and patterned without poling.

**Introduction**

Dramatic recent advancements in organic and molecular-scale electronic devices have yielded tremendous promise for lightweight, flexible devices with broad applications ranging including efficient lighting and displays[1-2] to solar cells[3-6], soft touch sensors,[7] and even magneto-optical sensors.[8] Despite such success, further fundamental research into integrated efficient energy storage, conversion, and generation mechanisms are critical.

Recent investigations in inorganic piezoelectric nanostructured ZnO and related materials have shown promise to interconvert mechanical and electrical energy for piezoelectric fabrics,[9] nanogenerators powered by sound waves,[10] and self-powered displays and sensors.[11] For organic and biological materials, bulk piezoelectric response has been measured in semicrystalline polymers like polyvinylidene difluoride (PVDF),[12] polar organic crystals,[13] polymer foams,[14] and even skin,[15] yet only recently has nanoscale characterization been possible. Modern atomic force microscopy (AFM) and piezoresponse force microscopy (PFM)[16] now enable us to probe the limits of piezoelectric distortion, including the piezoresponse of biological materials, such as individual collagen fibrils,[17] blood cells,[18] peptide nanotubes,[19] and viral capsids.[20]

Rather than a bulk response, our computational simulations have demonstrated that molecular clusters[21-22] and single molecules[23] can be highly piezoelectric, changing conformation in response to an applied electric field. That is, a molecular "spring" would extend and contract dependent on the direction and magnitude of the field (Fig. 1A). An obvious choice is a short oligopeptide with helical character, while rigid alkanes should show no conformational change along the molecular axis in response to the electric field, and can therefore serve as controls.

In this work, we experimentally demonstrate the converse piezoelectric deformation, causing a mechanical distortion of the preferred molecular conformation across single molecular monolayers of oligopeptides using an applied electric field. The piezoelectric response of the oligopeptides is markedly greater than control experiments with rigid alkanethiols and correlates with our combined theoretical predictions of conformational change driven by applied electric fields and surface spectroscopy. These results suggest that the piezoelectric response of biological materials is due to conformational changes in aligned helical domains. Unlike existing piezoelectric oxides, these peptide monolayers are intrinsically flexible, easily fabricated, aligned and patterned without poling, and possess strong piezoresponse on the nanoscale. Our results suggest that many proteins may exhibit a significant electromechanical response, since electrostatic fields due to ions or polar molecules such as water are ubiquitous on the nanoscale. We anticipate that a wide class of flexible polar molecules are piezoelectric, and a new generation of energy harvesting materials may be built up from intrinsic molecular conformational changes.

**Experimental and Computational Methods**

*Materials Used*

CA$_6$, A$_6$C, dodecanethiol (DT), and 11-mercaptoundecanethiol (MUA) compounds were purchased from Sigma-Aldrich, Inc. Amide-terminated CA$_6$ was purchased from AnaSpec, Inc. All molecules were used as received without further purification. The epoxy resin (Epo-Tek 377) was obtained from Epoxy Technology. Silicone elastomer kit (Sylgard 184) for making PDMS was from Dow Corning Corp. High grade mica substrate for template-stripped gold (TS-gold) was purchased from Ted Pella, Inc. Optical diffraction grating with 1μm grooves was obtained from Rainbow Symphony, Inc. The "dot" patterned master (2 $\mu$m patterned photoresist on silica wafer) for CA$_6$/MUA and MUA/CA$_6$-am mixed SAMs was obtained from the Center for Nanotechnology (CNT) Nanotechnology User Facility (NTUF) of the University of Washington.

*Sample Preparation*

*Template-stripped Gold (TS-gold)*

TS-gold[24] was prepared by depositing 200 nm of gold onto a freshly-cleaved mica surface via E-beam evaporation (Multi-source Electron Beam Evaporation System, Thermionics Laboratory VE180) with a base pressure of $2\times10^{-6}$ torr. The first 50 nm thick layer was deposited at a rate of 0.1 Å/s, while the remaining 150 nm was deposited at 0.5 Å/s. The epoxy adhesive starting materials were mixed according to the manufacturer's instruction. The gold side of the mica slice was then affixed to a glass slide with the epoxy precursor mixture and was placed in an oven at 150°C for 1 hour to anneal. The mica-gold-epoxy-glass sandwich was then immersed in tetrahydrofuran (THF) for several minutes to loosen the contact between the mica and gold film. The mica chips were then peeled off in THF using a pair of tweezers. The fresh gold surface was ready to use after drying with a stream of dry nitrogen.

*Patterned PDMS Stamp*

The two components of the silicon elastomer were mixed and vigorously stirred for 10 minutes and placed under vacuum for 30 min to drive out all bubbles. When clear, the mixture was poured into the mold (diffraction grating and "dot" masters), and then placed in a vacuum oven at 60°C for 1.5 hours. The cured polymer stamp was peeled off and cut into pieces approximately 2 mm × 4 mm. Before each stamping, the stamp was washed with acetone and isopropanol and dried with a gentle flow of nitrogen.

*Patterned and Mixed SAMs*

A solution of 10 μM of dodecanethiol (DT) in ethanol, 10 μM of 11-mercaptoundecanoic acid (MUA), ~1 mM of $CA_6$, ~1 mM of amidated-$CA_6$ ($CA_6$-am), and ~1 mM of $A_6C$ in 1:1 (v:v) water / acetonitrile were prepared as the "ink" source (due to the small amount of $CA_6$ and $A_6C$ used, an exact concentration of the oligoaminoacids was not determined). The stamp was dipped into 1 mL of the ink source for 5 minutes, dried in air and then stamped on the already prepared gold surface for 5 minutes before peeling off. Patterned DT, $CA_6$, and $A_6C$ SAMs were made following the same procedure, respectively. Patterned MUA SAMs were obtained using a drop cast method. A newly prepared stamp was used for each stamping.

$CA_6$ and MUA mixed films were obtained by immersing as-made patterned $CA_6$ films into 1 mM of MUA in ethanol for one hour, followed by washing with ethanol and drying in air. MUA and $CA_6$-am mixed films were obtained by immersing as made patterned MUA films into $CA_6$-am solution for one hour, followed by washing with water/acetonitrile and ethanol and drying in air.

*Piezoresponse Force Microscopy (PFM) Characterization*

PFM measurement was performed using an Asylum Research MFD-3D SPM with dual-AC resonance tracking (DART) PFM mode, unless otherwise specified in the text. Ti/Pt coated silicon tips (AC240TM, Asylum Research) with a first mode resonance frequency of 70 kHz and a normal stiffness of 2 N/m were used. For each sample characterized with DART, the contact resonance was determined, usually around 280 kHz, and then a 1-3 V AC bias was applied. Topography, piezoresponse amplitude, and phase images were all recorded. All reported amplitude and values have been corrected for the sample resonance enhancement (i.e., q-corrected) using the instrument default analyzing software. Multiple samples were fabricated and measured for consistency. For all quantitative measurements, inverse optical lever sensitivity (Invols)[25] was calibrated for every tip against a clean glass slide.

*FTIR Characterization*

*Sample Preparation*

Gold-coated quartz crystal electrodes (CH Instruments, Inc.) were used for the preparation of $CA_6$ (SAMs). The electrode was cleaned by soaking in piranha solution (concentrated sulfuric acid to hydrogen peroxide is 3:1) for two hours. The gold electrode was then immersed in $CA_6$ solution (as mentioned above) for 48 hours to insure complete formation of compact $CA_6$ SAMs, followed by washing with water/ethanol and drying in air. The SAMs coated gold electrode was then connected to one electrode of an alkaline battery (± 9.6 V) (and/or two batteries in series connection, +19.2 V), to charge the surface when performing FTIR grazing angle measurements. The charged sample was thus biased against ground, and an exact electric field across the monolayers cannot be determined. A gold electrode treated with piranha solution was used as background.

*Characterization*

Fourier Transform Infrared Microscope (FTIR) (Bruker VERTEX-70LS FTIR and Hyperion 2000) was used with the grazing angle mode. For a typical measurement, 0.41 to 1 $cm^{-1}$ resolution was chosen for a scan of 5000 times within the range from 1000 $cm^{-1}$ to 4000 $cm^{-1}$. The experiment was performed in ambient conditions, and the raw data was corrected for $CO_2$ and $H_2O$ effect by the default OPUS software.

*Computational Methods*

*Molecular Dynamics:*

The molecular dynamics simulations consisted of a 7x7 array of 49 peptides, with sequences CAAAAAA ($CA_6$) or AAAAAAC ($A_6C$), arranged on a plane that represented a hexagonal gold surface[26] with Au…Au spacing of 2.88 Å. Each peptide started as ideal α-helix and the sulfur atom of each cysteine in the peptide was held fixed. The initial coordinates of the peptides is included in the supplementary material and were generated using MOE[27]. For all MD runs, the Amber 99 force field parameters, as implemented in MOE, was used.[28]

*Molecular Dynamics Protocols*: The molecular dynamics simulations were performed using NAMD 2.8.[29] All simulations were done *in vacuo* using a 2 fs time step. Electrostatic and van der Waals interactions were switched off over that range of 2.5 Å starting at 10 Å. A pairlist for these interactions was calculated using a 13.5 Å distance. A temperature of 300 K was used in the simulations and was kept constant using a Langevin thermostat. The simulations were run for a total of 1 ns. For the non-zero field simulation, a 2.306 kcal/mol•Å•e (= 1.0 V/nm) or 10.0 kcal/mol•Å•e (=4.35 V/nm) external electric field was applied in the +z direction, normal to the planar surface defined by the sulfur atoms of the peptides.

The trajectory file from each of the different simulations was analyzed using VMD.[30] First, the average height of each of peptide above the "gold" surface was calculated. The average height for each peptide was obtained by averaging the z coordinate for the terminal nitrogen atom of over frames 2500-4500 of a 5000 frame trajectory. Secondly, the percent helix character for each peptide was calculated, over the same 2000 frames used to determine the average height above the surface. The helix content was determined using the "sscalc" routine in VMD. The output of the "sscalc" routine was used to determine the percent helix by adding the number of times a residue was determined to be an "h" (α-helix) or "g" ($3_{10}$ helix) and dividing that result by 7 (the total number of residues).

*Density Functional Theory Geometry Optimizations:*

We used Gaussian 09[31] and density functional theory (DFT), with the B3LYP functional[32-33] and the 6-31G(d) basis set to optimize all computed structures, including optimization of molecule length, dipole moment, and energy under different applied electric bias, as performed in our previous work. To consider the conformational change in response to the electric field, the molecule was oriented to a specific frame of reference using Avogadro[34] and the specific direction and magnitude of the field ±1.29 V/nm (±25×10$^{-4}$ a.u.) was added to the Gaussian input along the z-axis, defined as along the molecule helix ($CA_6$, $CA_6$-am, and $A_6C$) and the carbon chain (DT and MUA). The molecule height is defined as the distance between two atoms at the far ends of each molecule and is consistent throughout all measurements.

To compute the piezo coefficient ($d_{33}$), the following unit conversion was performed:

$$d_{33}(pm/V) = \frac{\Delta z(\text{Å})}{z(\text{Å})F(V/nm)} \times \frac{10^3 pm}{1 nm}$$

where z is the height of the optimized geometry at zero electric field, Δz is the difference in height between the optimized geometry at a minor field strength (e.g. 0.257 V/nm) and at zero electric field, and F is the corresponding field range. In short, the response is the fractional length change per unit of applied electric field.

*Molecular Length of $CA_6$, $A_6C$, $CA_6$-am, DT, and MUA:*

The molecular length was defined as the distance between the two atoms on the furthest ends of the optimized molecule plus van der Waals radius[35] of the two atoms chosen. The molecular length is thus 15.0 Å for $CA_6$, 16.8 Å for $A_6C$, 17.1 Å for $CA_6$-am, 19.1 Å for DT, and 18.8 Å for MUA.

**Results and Discussion**

In this experiment, we have created patterned single monolayer piezoelectrics with intrinsic polar ordering. A solution of short oligopeptides (e.g., with sequence CAAAAAA, or $CA_6$ Fig. 1B) was prepared as the "ink source" for microcontact patterning[36] to smooth template-stripped gold substrates[24] through the cysteine-gold interaction. The patterning with ~1 $\mu$m spacing between lines yields intentionally incomplete coverage of the oligopeptides across the gold surface, so that only a single monolayer forms via self-assembly. Compared to previous studies based on surface grafting,[37] microcontact printing and solution-phase self-assembly provides an easier way to form patterned films.

To determine the piezoelectric response, we used PFM with dual AC resonance tracking (DART) mode,[38] where the tip is in mechanical contact resonance with the oligopeptide surface, a sinusoidal bias voltage is applied, and the corresponding height amplitude is determined. AFM topography (Fig. 1C, S1) indicates a film height (~1.5 nm) that corresponds closely to the computed molecular length (1.50 nm, Supporting Information). Piezo amplitude of $CA_6$ monolayers shows ~ 6 pm under 3V bias from scan lines, yielding a piezoelectric coefficient of ~2 pm/V (Fig. 1D, S2). While the AC bias is in the range of 1-3V, the exact field across the monolayer is much less, as discussed below. Moreover, samples are scanned repeatedly, ensuring no sample degradation or decomposition is observed. Results are consistent across multiple scans and averaged over multiple samples.

DART PFM is believed to be an improved technique for measuring low response and non-uniform piezoelectric materials while providing a lower signal-to-noise ratio and fewer resonance artifacts. There remain some limitations with obtaining a quantitative PFM measurement for weak and soft piezoelectric materials, including nonlocal effects,[39] complex background signal,[40] tip-surface electrostatic effects,[41] and potential drop from tip to surface due to weak indentation and high surface resistance.[41]

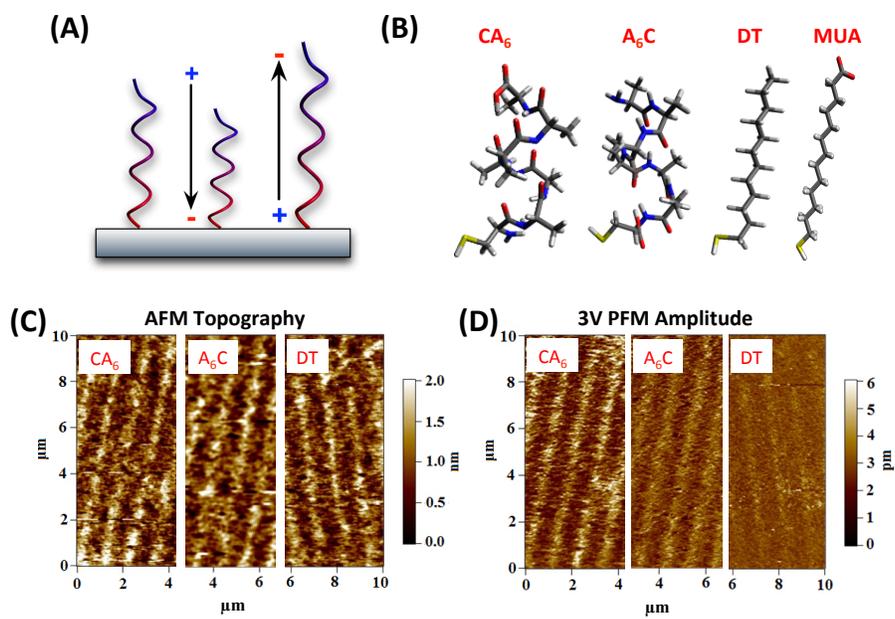

**Figure 1.** Schematic of molecular springs and realization through AFM topography and DART PFM amplitude of patterned molecular monolayers. (A) Schematic of piezoelectric distortion through compression and extension of a molecular helix, (B) Molecular structures of oligopeptides $CA_6$, $A_6C$, and rigid alkanes DT and MUA, (C) AFM topography of oligopeptide $CA_6$ (left), oligopeptide $A_6C$ (middle), and DT, (right) patterned monolayers, and (D) DART PFM amplitude at an applied bias of 3V for patterned $CA_6$, $A_6C$, and DT monolayers, showing contrast with bare gold.

To overcome these artificial amplitude contributions, we introduced built-in controls to examine the absolute piezoresponse. Patterned single monolayers of another oligopeptide (AAAAAAC, $A_6C$) and dodecanethiol (DT) were deposited by the same method onto individual gold substrates (Fig. 1C), to form single monolayers, again judging by the agreement between AFM topography and computed molecular size.[42] Significant piezo response is found via DART PFM for both $CA_6$ and $A_6C$ stripes and substantially less amplitude is observed from the DT monolayers (Fig. 1D, S2). Thus, we hypothesize that the relatively large PFM amplitude from the $CA_6$ and $A_6C$ monolayers derives primarily from intrinsic molecular conformational changes caused by the applied electric field. Some of the observed difference in PFM amplitude between $CA_6$ and $A_6C$ monolayers may derive from the terminal (top) end group effect, -COOH for $CA_6$ and $-NH_2$ for $A_6C$, respectively.

Since the molecular monolayer (~1.5 nm thickness) between the tip and the gold substrate forms a capacitance layer under applied AC bias, the electrostatic effect has to be ruled out in explaining the amplitude contrast of the observed piezoresponse between $CA_6$ and DT (and between $CA_6$ and MUA as well as shown later). Two major aspects should be taken into account to determine the magnitude of the electrostatic effect, the dielectric constant $\varepsilon$ and the Young's modulus $E$ of the monolayer. A larger dielectric constant of the film underneath the tip is directly proportional to the electrostatic force,[43-44] while a larger modulus yields smaller deformation for the same applied force. In our experiment, the electrostatic contribution of the measured piezoresponse will depend on the relative $\varepsilon/E$ ratios of $CA_6$ and DT. The dielectric constant for alkanethiol SAMs is in the range of 2.0 to 2.7 measured via different methods,[45-47] while the dielectric constant for non-solvent proteins is estimated to be somewhat larger, 2.4-3.9 for polyalanine.[48] However, the peptide monolayer is reported to have a modulus larger than 10 GPa,[37] more than 5-fold stiffer than its alkanethiol counterpart of 2 GPa and below.[49-51] The electrostatic force should then induce a larger deformation on the DT SAMs, rather than the larger piezoresponse observed of $CA_6$ SAMs.

Since the PFM characterization has been performed in ambient conditions, a water meniscus will form at the surface-tip interface, decreasing the effective electric field across the SAM. At neutral pH, the C-terminus of $CA_6$ will partially yield an anionic carboxylate-terminated surface atop the oligopeptide monolayer (vide infra), while the alkyl-terminated DT remains neutral. Similarly, $CA_6$ shows greater PFM response than the amine-terminated $A_6C$. Such electrostatic response would not only be different between the bare gold surface and the carboxylate-terminated $CA_6$ monolayers, but may explain the larger response of $CA_6$ when compared to the nonpolar, alkyl-terminated DT monolayers. To control for the possible surface charge difference, we prepared films using $CA_6$ peptide monolayer "dot" patterns (Fig. 2A), followed by back-filling the remaining exposed gold surface with 11-mercaptoundecanoic acid (MUA), since both molecules have a carboxylic acid group at the surface end (Fig. 2C). The isoelectric point (pI) of $CA_6$ is estimated to be 5.5, close to the pKa ~4.8 of MUA in solution.[52] The similarity between these two ends is also demonstrated by surface potential measurement via Kelvin probe force microscopy (KPFM), which gives a much smaller surface potential difference <20 mV between MUA and $CA_6$ compared to >70 mV between $CA_6$ and gold (Supporting Information, Fig. S3).

Both $CA_6$ and MUA have similar computed heights (1.50 nm and 1.88 nm, respectively, Supporting Information) so the AFM topology of the mixed film (Fig. 2C) appears essentially featureless; although the phase channel (Fig. 2E) shows slight variation between regions of the two molecular monolayers. While both $CA_6$ and MUA have comparable heights and terminal carboxylic acid groups, the computed piezo-driven conformational change in MUA (vide infra) is dramatically smaller than $CA_6$. As seen experimentally in Fig. 2G, the $CA_6$ islands exhibit markedly greater PFM amplitude at 3V bias than MUA. These mixed $CA_6$/MUA films are highly stable, and we have observed pattern retention and strong PFM response over 42 days in ambient conditions (Fig. S4).

To further demonstrate that the observed piezoelectric deformation is generated by conformational changes, not the electrostatic effect of the surface end groups, mixed films of charged MUA and amide-terminated $CA_6$ ($CA_6$-am) were tested. In $CA_6$-am the terminal carboxylate group is converted to $-CONH_2$, so the surface end will remain electrically neutral. Fig. 2B shows the dot patterned MUA SAMs and Fig 2D shows the topographically-featureless MUA/$CA_6$-am mixed films since the molecules have comparable heights. The phase image (Fig. 2F) shows a stronger contrast between MUA and $CA_6$-am domains indicating a greater surface property differentiation of the $-COO^-$ and $-CONH_2$ groups. The surface neutral $CA_6$-am SAMs still provides significantly larger piezo amplitude than MUA SAMs. While the single-frequency PFM mode used to image the two mixed films cannot yield reliable *quantitative* measurements, this technique clearly indicates that PFM deformation observed in $CA_6$ and $CA_6$-am patterns derives largely from molecular conformational changes and from end-group effects. Such piezo amplitude contrast observed on a uniform surface rules out the problematic coincidence of topography and piezoresponse.[53]

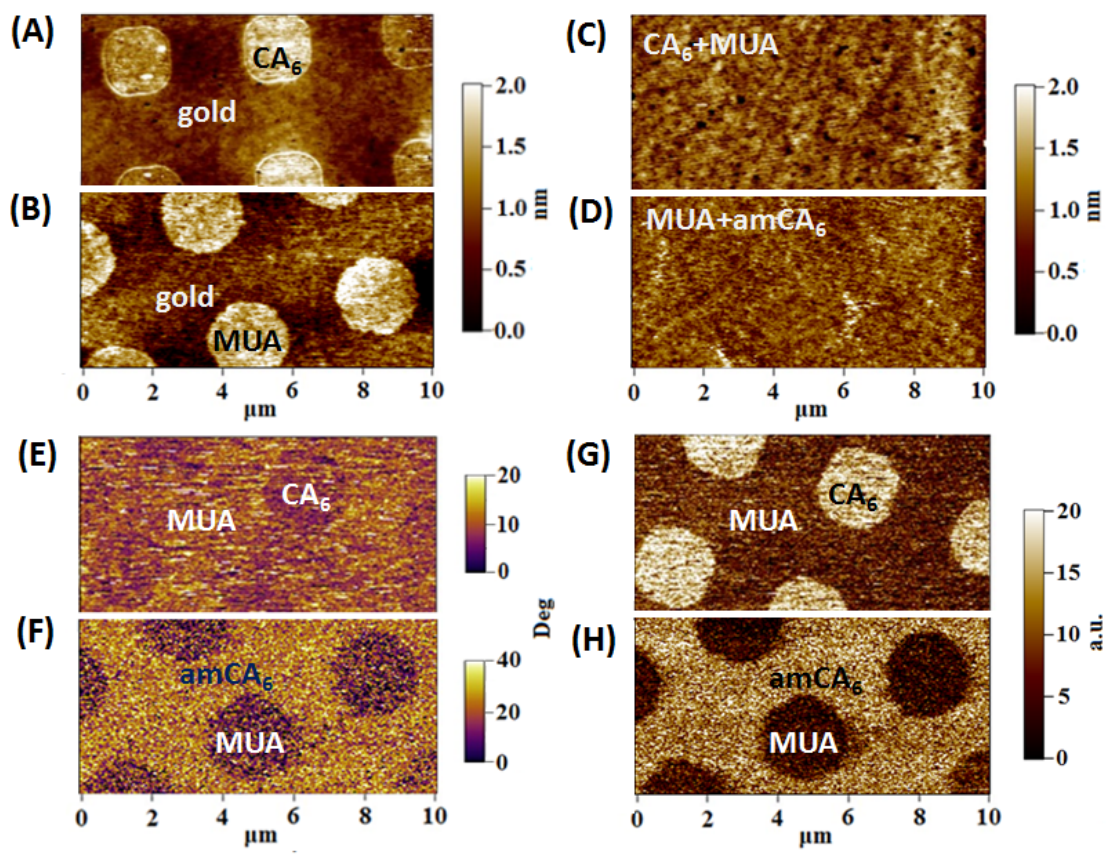

**Figure 2.** AFM topography, single-frequency PFM phase and amplitude of patterned and mixed monolayers. AFM topography of oligopeptide $CA_6$ SAM (A) and MUA SAM (B) on bare gold, AFM topography of the mixed SAMs $CA_6$/MUA (C) and MUA/$CA_6$-am (D), single-frequency PFM phase image of mixed $CA_6$/MUA (E) MUA/$CA_6$-am (F), and single-frequency PFM amplitude of a mixed $CA_6$/MUA (G) and MUA/$CA_6$-am (H) films with an applied bias of 3V.

Grazing-angle Fourier transform infrared spectroscopy (GA-FTIR) was used to examine the conformational change of oligopeptide $CA_6$ under different applied electric fields. The so-called amide I peak (1652-1657 cm$^{-1}$) and amide II peak (1545-1551 cm$^{-1}$), which are absorptions from combinations of C=O bond stretching, C-N bond stretching, and N-H bond bending, are indicative of the secondary structures of proteins.[54] As shown in Fig. S5, clear peak shift is observed when the gold substrate of $CA_6$ SAMs was charged with different potentials using an external battery. Although the broad peak widths indicate a highly diverse ensemble, making exact assignment of the vibrational peaks difficult, these results are consistent with previous studies.[55]

While results from DART PFM, mixed films, and FTIR strongly suggest an electromechanical distortion indicating from conformational change in the molecular monolayer; to be piezoelectric, this distortion should be linear in response to applied bias. As the PFM drive bias is increased, the measured amplitude increases substantially (Fig. 3A). The PFM amplitude, however, is averaged across an ensemble of ~$10^4$ molecules, based on the effective tip radius and convoluted with nonlocal and electrostatic response, not just the piezoelectric deformation of the monolayers. Consequently, to quantitatively consider the response of the $CA_6$ monolayers to the electric field, we compiled histograms of the PFM amplitudes as a function of electric field in DART mode (Fig. 3B, S6). The low-response (left) edge of the histogram serves as a baseline for the gold surface, and the high-response (right) edge of the histogram indicates the molecular extension. Both line scans and histograms indicate a large number of molecules exhibit large responses. For quantitative analysis, we used the full width at half-maximum (FWHM) as a measure of the PFM amplitude. The results averaged across multiple regions on multiple films measured with multiple tips indicate a linear FWHM change in response to applied electric field (Fig. 3C).

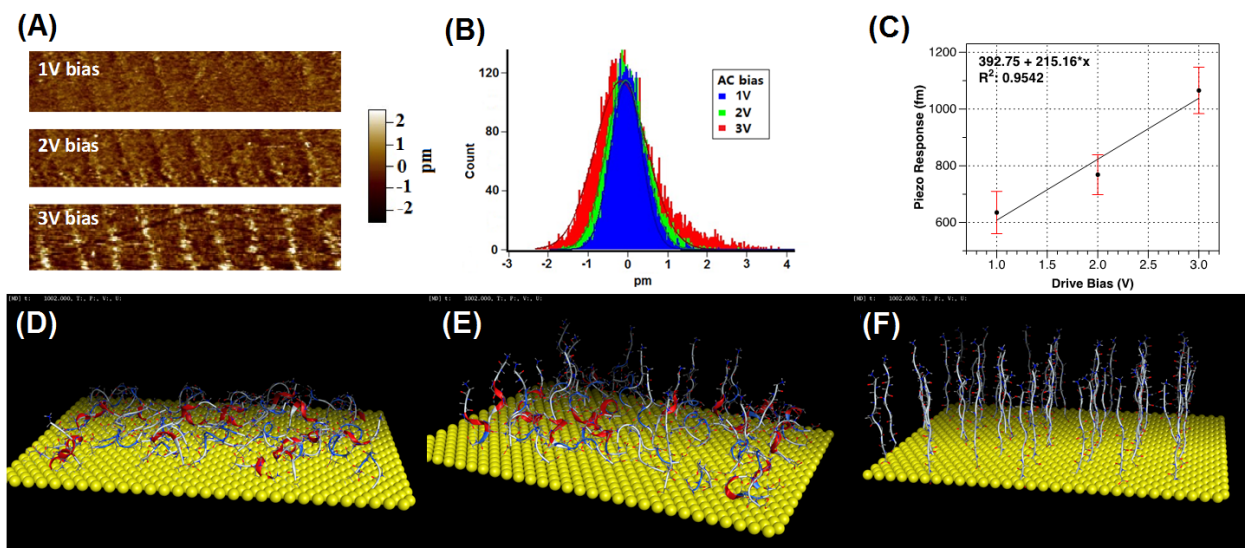

**Figure 3.** Piezo amplitude of $CA_6$ SAMs as a function of increasing applied bias voltage. (A) DART PFM amplitude of $CA_6$ SAMs on bare gold, showing clearly increasing amplitude with increasing bias, (B) example histograms of DART PFM amplitudes as a function of increasing bias voltage, and (C) linear correlation between FWHM of histograms (as a measure of "average" response) and applied drive bias voltage. (D-F) Molecular dynamics snapshots of a simulated $CA_6$ monolayer as a function of increasing electric field at (D) 0 V/nm, (E) 1 V/nm, and (F) 4.3 V/nm.

Since the $CA_6$ sequence is short, despite the presence of an oligo-alanine block, the preferred conformation in solution is likely to be a random coil. The predominant conformation, however, has been shown to change due to assembly on metal surfaces, particularly in the presence of electric field gradients.[55]

To address the preferred conformation of $CA_6$ in our system, particularly in light of the surface FTIR evidence, molecular dynamics (MD) calculations were performed on an array of $CA_6$ molecules, simulating the self-assembled monolayers, both with and without an applied electric field. Based on the pI ~5.5 of $CA_6$,[21] most molecules in the monolayer will be anionic at the carboxylate terminus, but the amine terminus is neutral and unprotonated. Computed MD trajectories using zero applied electric field or fields smaller than 1 V/nm are predicted to exhibit no helical structure or insignificant molecular extension (Fig. 3D). Upon increasing the applied electric field to 0.5-1 V/nm, the $CA_6$ molecules extend in the applied field and adopt more helical or linear extended conformations (Fig. 3E, S7). At an applied field of 4.3 V/nm, the MD trajectories predict the $CA_6$ molecules adopt almost exclusively a linear extended conformation (Fig. 3F), with a correspondingly large increase in average extension above the surface.

While these MD simulations confirm the experimental observation of piezoelectric deformation in $CA_6$ monolayers, the predicted piezoresponse is much too large (~7.5Å, Fig. S7), compared with experiment. Furthermore, classical force fields used in MD cannot properly treat the polarizable electrostatic response of these molecules to an applied electric field. Consequently, full geometry optimizations using density functional theory (DFT) methods were performed as a function of applied electric field (*vide infra*). As shown in Fig. S8A-B, the length of DT and MUA are not predicted to change substantially, but helical $CA_6$ exhibits a large deformation along the molecular axis. Bond lengths do not distort appreciably; instead the applied field alters the pitch of the helix. This agrees with previous studies which show that the structural origin of protein/polypeptide piezoelectric materials lies in the orderly aligned secondary structure including α-helix[56] and β-sheet.[57]

The sharp length change of $CA_6$, predicted at ~1.1 V/nm, involves a secondary structure transformation from an α-helix to a $3_{10}$-conformation.[58] Such an abrupt conformational change might be expected because the applied field distorts the backbone dihedral angles and hydrogen bonding. Using the linear response around zero field strength (±0.26 V/nm), the predicted piezoelectric coefficient ($d_{33}$) of $CA_6$ is computed to be 14.7 pm/V, comparable to ZnO ($d_{33}$ ~9.9-27 pm/V)[59] and PVDF ($d_{33}$ -26 pm/V).[60] Consistent with the large experimentally observed difference in PFM amplitudes between $CA_6$, DT, and MUA (e.g., ~10 a.u. for $CA_6$ / MUA in Fig. 2D), computed piezo coefficients for DT and MUA were small (0.54 pm/V and 1.55 pm/V for DT and MUA, respectively).

Fully extended, linear conformations of $CA_6$ are also predicted to only exhibit small piezo coefficients (1.01 pm/V, Fig. S8C), similar in magnitude to MUA. Thus, these conformations would not be expected to show the greater PFM response in the mixed $CA_6$/MUA films (Fig. 2D). When random coil conformations, selected from the MD trajectories at low electric field strength, are optimized as a function of applied electric field, distortions of conformation do occur but not as a smooth linear monotonic function of electric field strength (Fig. S8D). This confirms the MD results — namely that random or linear conformations of the $CA_6$ oligopeptide would not reliably, repeatedly deform with linear response to an applied electric field as observed experimentally.

Qualitatively, both DFT-computed response and PFM characterization yield the same conclusion, namely that $CA_6$ exhibits substantial piezo deformation, while DT and MUA do not because of their rigid molecular shape. The quantitative comparison of computed and PFM-measured $d_{33}$ piezo coefficients for $CA_6$ is good (14.7 pm/V for DFT and up to ~2 pm/V from PFM, based on line scans). The DFT-computed piezoresponse of completely helical $CA_6$ represents an ideal upper bound of the experimentally observed deformation, since the calculation involves a field applied exactly parallel to the molecular axis to a peptide in vacuum and assumes perfect electromechanical coupling. Moreover, peptides with some fraction of both helical and linear conformation, as suggested by MD would have an average of the two responses (e.g., ~7 pm/V for a 50:50 mixture). The PFM technique is, at best, semi-quantitative, since the field from the conductive tip will be applied through an aqueous meniscus, and across a time-varying distance, due to the mechanical frequency of the tapping. Lower indentation on soft materials also provides an effective piezoelectric coefficient smaller than the actual value.[41] Furthermore, based on the tip resolution, the PFM measurement will sample an ensemble of ~$10^4$ molecules which may not be responding coherently to the AC voltage. For example, recent studies have confirmed that SAMs of helical molecules are disordered and exhibit varying tilt angles near the periphery.[61]

**Conclusions**

Our results indicate a repeatable, continuous, linear electromechanical response of oligopeptides to applied nanoscale electric fields, indicating a molecular converse piezoelectric effect, predicted by computational studies.[23] We note that many previous studies have demonstrated the piezoactivity of biological materials, and we find that our measured piezoelectric response of monolayers of short peptides correspond well with such reports (in the order of 1 pm/V). While the applied AC bias voltages used in the PFM experiment appear large (e.g., 3 V), the effective field across the monolayer is undoubtedly much smaller due to screening from interfacial meniscus between tip and surface and judging by the larger computed piezoresponse using DFT. Moreover, a field of ~1 V/nm corresponds to that generated by a singly-charged ion at distance of 1 nm from a molecule. Consequently, fields of this magnitude, while large on a bulk scale, are common on the molecular scale.

We speculate that many proteins may exhibit significant electromechanical conformational response, since such electrostatic fields due to molecular dipole moments, ions, etc. are ubiquitous on the nanoscale. Our results also suggest that the observed piezoactivity of skin, muscle, and other biomaterials, discussed earlier, may result from aligned helical, polar domains.

There are several key implications of our results. First, it suggests that piezoelectric energy conversion can be used to self-power nanoscale organic electronics, such as flexible touch sensors. Since the piezoelectric effect also converts electric fields to motion, it also can be used to generate reliable nanoscale linear movement. Existing inorganic nanopiezotronics such as ZnO nanoribbons are also difficult to align and pattern, while we demonstrate that simple self-assembly and solution patterning work with molecular piezoelectrics. Finally, one can imagine that the diversity of chemical synthetic techniques, combined with computational simulation and PFM characterization can be used to design highly piezo-responsive monolayers. Such molecules must retain polarity, but need not be chiral or asymmetric. If possible, this would enable a new class of piezomaterials, in which the deformation derives from intrinsic molecular conformational changes.

**Acknowledgments** We thank Dr. Sergei Kalinin, Dr. Keith Jones, Dr. Roger Prokosh, and Prof. David Waldeck for helpful suggestions. Thanks also to Prof. Lee Y. Park for loaning microcontact masters. We thank the University of Pittsburgh, ACS PRF (49002-DN110), AFOSR (FA9550-12-1-0228), and RCSA through the Cottrell Scholar Award for funding. Data reported in this work can be found in the Supplementary Material.

**Author Information**

**Contributions** X.Q. and G.R.H. designed the experiments and simulations; J.D.M. performed the molecular dynamics simulations; X.Q. performed the DFT simulations and experiments; and G.R.H. and X.Q. analyzed the data and wrote the paper. All authors discussed the results and commented on the manuscript.

**Competing financial interests:** The authors declare no competing financial interests.

**Supporting Information**. AFM topology, Kelvin-probe microscopy, DART PFM scans, surface FTIR, molecular dynamics analysis, and DFT calculation analysis, are included as supporting information. This material is available free of charge via the Internet at http://pubs.acs.org

**References and Notes**

1. Gather, M. C.; Kohnen, A.; Meerholz, K., White organic light-emitting diodes. *Adv Mater* **2011,** *23* (2), 233-48.
2. Ostroverkhova, O., Organic Optoelectronic Materials: Mechanisms and Applications. *Chem Rev* **2016,** *116* (22), 13279-13412.
3. He, F.; Yu, L., How Far Can Polymer Solar Cells Go? In Need of a Synergistic Approach. *J Phys Chem Lett* **2011,** *2* (24), 3102-3113.
4. Zhao, W.; Li, S.; Yao, H.; Zhang, S.; Zhang, Y.; Yang, B.; Hou, J., Molecular Optimization Enables over 13% Efficiency in Organic Solar Cells. *J Am Chem Soc* **2017,** *139* (21), 7148-7151.

5. Cui, Y.; Yao, H.; Gao, B.; Qin, Y.; Zhang, S.; Yang, B.; He, C.; Xu, B.; Hou, J., Fine-Tuned Photoactive and Interconnection Layers for Achieving over 13% Efficiency in a Fullerene-Free Tandem Organic Solar Cell. *J Am Chem Soc* **2017,** *139* (21), 7302-7309.
6. Bin, H.; Yang, Y.; Zhang, Z. G.; Ye, L.; Ghasemi, M.; Chen, S.; Zhang, Y.; Zhang, C.; Sun, C.; Xue, L.; Yang, C.; Ade, H.; Li, Y., 9.73% Efficiency Nonfullerene All Organic Small Molecule Solar Cells with Absorption-Complementary Donor and Acceptor. *J Am Chem Soc* **2017,** *139* (14), 5085–5094.
7. Chen, D.; Pei, Q., Electronic Muscles and Skins: A Review of Soft Sensors and Actuators. *Chem Rev* **2017,** *117* (17), 11239-11268.
8. Swager, T. M., 50th Anniversary Perspective: Conducting/Semiconducting Conjugated Polymers. A Personal Perspective on the Past and the Future. *Macromolecules* **2017,** *50* (13), 4867-4886.
9. Wang, Z. L.; Song, J., Piezoelectric nanogenerators based on zinc oxide nanowire arrays. *Science* **2006,** *312* (5771), 242-6.
10. Cha, S. N.; Seo, J. S.; Kim, S. M.; Kim, H. J.; Park, Y. J.; Kim, S. W.; Kim, J. M., Sound-driven piezoelectric nanowire-based nanogenerators. *Adv Mater* **2010,** *22* (42), 4726-30.
11. Xu, S.; Qin, Y.; Xu, C.; Wei, Y.; Yang, R.; Wang, Z. L., Self-powered nanowire devices. *Nat Nanotechnol* **2010,** *5* (5), 366-73.
12. Fukada, E.; Takashita, S., Piezoelectric Effect in Polarized Poly (Vinylidene Fluoride). *Jpn J Appl Phys* **1969,** *8* (7), 960.
13. Horiuchi, S.; Tokura, Y., Organic ferroelectrics. *Nat Mater* **2008,** *7* (5), 357-66.
14. Moody, M. J.; Marvin, C. W.; Hutchison, G. R., Molecularly-doped polyurethane foams with massive piezoelectric response. *J Mater Chem C* **2016,** *4* (20), 4387-4392.
15. Athenstaedt, H.; Claussen, H.; Schaper, D., Epidermis of human skin: pyroelectric and piezoelectric sensor layer. *Science* **1982,** *216* (4549), 1018-20.
16. Kalinin, S. V.; Bonnell, D. A.; Alvarez, T.; Lei, X.; Hu, Z.; Ferris, J. H.; Zhang, Q.; Dunn, S., Atomic polarization and local reactivity on ferroelectric surfaces: A new route toward complex nanostructures. *Nano Lett* **2002,** *2* (6), 589-593.
17. Minary-Jolandan, M.; Yu, M. F., Nanoscale characterization of isolated individual type I collagen fibrils: polarization and piezoelectricity. *Nanotechnology* **2009,** *20* (8), 085706.
18. Raman, A.; Trigueros, S.; Cartagena, A.; Stevenson, A. P.; Susilo, M.; Nauman, E.; Contera, S. A., Mapping nanomechanical properties of live cells using multi-harmonic atomic force microscopy. *Nat Nanotechnol* **2011,** *6* (12), 809-14.
19. Kholkin, A.; Amdursky, N.; Bdikin, I.; Gazit, E.; Rosenman, G., Strong piezoelectricity in bioinspired peptide nanotubes. *ACS Nano* **2010,** *4* (2), 610-4.
20. Lee, B. Y.; Zhang, J.; Zueger, C.; Chung, W. J.; Yoo, S. Y.; Wang, E.; Meyer, J.; Ramesh, R.; Lee, S. W., Virus-based piezoelectric energy generation. *Nat Nanotechnol* **2012,** *7* (6), 351-6.
21. Werling, K. A.; Griffin, M.; Hutchison, G. R.; Lambrecht, D. S., Piezoelectric hydrogen bonding: computational screening for a design rationale. *J Phys Chem A* **2014,** *118* (35), 7404-10.
22. Werling, K. A.; Hutchison, G. R.; Lambrecht, D. S., Piezoelectric Effects of Applied Electric Fields on Hydrogen-Bond Interactions: First-Principles Electronic Structure Investigation of Weak Electrostatic Interactions. *J Phys Chem Lett* **2013,** *4* (9), 1365-70.

23. Quan, X. F.; Marvin, C. W.; Seebald, L.; Hutchison, G. R., Single-Molecule Piezoelectric Deformation: Rational Design from First-Principles Calculations. *J Phys Chem C* **2013,** *117* (33), 16783-16790.
24. Hegner, M.; Wagner, P.; Semenza, G., Ultralarge Atomically Flat Template-Stripped Au Surfaces for Scanning Probe Microscopy. *Surf Sci* **1993,** *291* (1-2), 39-46.
25. Alexander, S.; Hellemans, L.; Marti, O.; Schneir, J.; Elings, V.; Hansma, P. K.; Longmire, M.; Gurley, J., An Atomic-Resolution Atomic-Force Microscope Implemented Using an Optical-Lever. *J Appl Phys* **1989,** *65* (1), 164-167.
26. make_gold_surface.svl, Scientific Vector Language (SVL) source code provided by Chris Williams at Chemical Computing Group Inc., 1010 Sherbooke St. West, Suite #910, Montreal, QC, Canada, H3A 2R7, 2011
27. Molecular Operating Environment (MOE), 2011.10; Chemical Computing Group Inc., 1010 Sherbooke St. West, Suite #910, Montreal, QC, Canada, H3A 2R7, 2011.
28. Cieplak, P.; Caldwell, J.; Kollman, P., Molecular mechanical models for organic and biological systems going beyond the atom centered two body additive approximation: Aqueous solution free energies of methanol and N-methyl acetamide, nucleic acid base, and amide hydrogen bonding and chloroform/water partition coefficients of the nucleic acid bases. *J Comp Chem* **2001,** *22* (10), 1048-1057.
29. Phillips, J. C.; Braun, R.; Wang, W.; Gumbart, J.; Tajkhorshid, E.; Villa, E.; Chipot, C.; Skeel, R. D.; Kale, L.; Schulten, K., Scalable molecular dynamics with NAMD. *J Comput Chem* **2005,** *26* (16), 1781-802.
30. Humphrey, W.; Dalke, A.; Schulten, K., VMD: visual molecular dynamics. *J Mol Graph* **1996,** *14* (1), 33-8, 27-8.
31. Frisch, M. J.; Trucks, G. W.; Schlegel, H. B.; Scuseria, G. E.; Robb, M. A.; Cheeseman, J. R.; Scalmani, G.; Barone, V.; Mennucci, B.; Petersson, G. A. et al. Gaussian 09, Revision A.02, Gaussian, Inc. Wallingford, CT, USA, 2009
32. Becke, A. D., Density-Functional Thermochemistry .3. The Role of Exact Exchange. *J Chem Phys* **1993,** *98* (7), 5648-5652.
33. Lee, C. T.; Yang, W. T.; Parr, R. G., Development of the Colle-Salvetti Correlation-Energy Formula into a Functional of the Electron-Density. *Phys Rev B* **1988,** *37* (2), 785-789.
34. Hanwell, M. D.; Curtis, D. E.; Lonie, D. C.; Vandermeersch, T.; Zurek, E.; Hutchison, G. R., Avogadro: an advanced semantic chemical editor, visualization, and analysis platform. *J Cheminform* **2012,** *4* (1), 17.
35. Bondi, A., Van Der Waals Volumes and Radii. *J. Phys. Chem.* **1964,** *68* (3), 441-451.
36. Kumar, A.; Whitesides, G. M., Patterned condensation figures as optical diffraction gratings. *Science* **1994,** *263* (5143), 60-2.
37. Jaworek, T.; Neher, D.; Wegner, G.; Wieringa, R. H.; Schouten, A. J., Electromechanical properties of an ultrathin layer of directionally aligned helical polypeptides. *Science* **1998,** *279* (5347), 57-60.
38. Rodriguez, B. J.; Callahan, C.; Kalinin, S. V.; Proksch, R., Dual-frequency resonance-tracking atomic force microscopy. *Nanotechnology* **2007,** *18* (47), 475504.
39. Hong, J. W.; Park, S. I.; Khim, Z. G., Measurement of hardness, surface potential, and charge distribution with dynamic contact mode electrostatic force microscope. *Rev Sci Instrum* **1999,** *70* (3), 1735-1739.
40. Jungk, T.; Hoffmann, A.; Soergel, E., Quantitative analysis of ferroelectric domain imaging with piezoresponse force microscopy. *Appl Phys Lett* **2006,** *89* (16) 163507.

41.	Kalinin, S. V.; Bonnell, D. A., Imaging mechanism of piezoresponse force microscopy of ferroelectric surfaces. *Phys Rev B* **2002,** *65* (12), 125408.
42.	These three molucules with similar computed molecular lengths ($CA_6$ 1.50 nm, $A_6C$ 1.68 nm, and DT 1.91 nm) all form SAMs on gold but with different terminal (top) end group, e.g. -COOH/-COO- for $CA_6$, $-NH_2$ for $A_6C$ and $-CH_3$ for DT.
43.	Sacha, G. M.; Sahagun, E.; Saenz, J. J., A method for calculating capacitances and electrostatic forces in atomic force microscopy. *J Appl Phys* **2007,** *101* (2) 024310.
44.	Belaidi, S.; Girard, P.; Leveque, G., Electrostatic forces acting on the tip in atomic force microscopy: Modelization and comparison with analytic expressions. *J Appl Phys* **1997,** *81* (3), 1023-1030.
45.	Rampi, M. A.; Schueller, O. J. A.; Whitesides, G. M., Alkanethiol self-assembled monolayers as the dielectric of capacitors with nanoscale thickness. *Appl Phys Lett* **1998,** *72* (14), 1781-1783.
46.	Damos, F. S.; Luz, R. C.; Kubota, L. T., Determination of thickness, dielectric constant of thiol films, and kinetics of adsorption using surface plasmon resonance. *Langmuir* **2005,** *21* (2), 602-9.
47.	Kumar, B.; Bonvallet, J. C.; Crittenden, S. R., Dielectric constants by multifrequency non-contact atomic force microscopy. *Nanotechnology* **2012,** *23* (2), 025707.
48.	Gilson, M. K.; Honig, B. H., The dielectric constant of a folded protein. *Biopolymers* **1986,** *25* (11), 2097-119.
49.	Salmeron, M.; Neubauer, G.; Folch, A.; Tomitori, M.; Ogletree, D. F.; Sautet, P., Viscoelastic and Electrical-Properties of Self-Assembled Monolayers on Au(111) Films. *Langmuir* **1993,** *9* (12), 3600-3611.
50.	Price, W. J.; Leigh, S. A.; Hsu, S. M.; Patten, T. E.; Liu, G. Y., Measuring the size dependence of Young's modulus using force modulation atomic force microscopy. *J Phys Chem A* **2006,** *110* (4), 1382-8.
51.	DelRio, F. W.; Jaye, C.; Fischer, D. A.; Cook, R. F., Elastic and adhesive properties of alkanethiol self-assembled monolayers on gold. *Appl Phys Lett* **2009,** *94* (13).
52.	Kozlowski L. 2007-2012 Isoelectric Point Calculator. http://isoelectric.ovh.org/. The pKa of MUA (or CA6 or any surface-bound acid) depends both on the local ionic activity and measuring methods. pKa of MUA in solution is however ~4.8, e.g. [D. Wang et al., "How and Why Nanoparticle's Curvature Regulates the Apparent pKa of the Coating Ligands." *J. Am. Chem. Soc.* **2011** 133, 2192.]
53.	Jesse, S.; Baddorf, A. P.; Kalinin, S. V., Dynamic behaviour in piezoresponse force microscopy. *Nanotechnology* **2006,** *17* (6), 1615-28.
54.	Jackson, M.; Mantsch, H. H., The use and misuse of FTIR spectroscopy in the determination of protein structure. *Crit Rev Biochem Mol Biol* **1995,** *30* (2), 95-120.
55.	Chen, Y.; Cruz-Chu, E. R.; Woodard, J. C.; Gartia, M. R.; Schulten, K.; Liu, L., Electrically induced conformational change of peptides on metallic nanosurfaces. *ACS Nano* **2012,** *6* (10), 8847-56.
56.	Farrar, D.; Ren, K.; Cheng, D.; Kim, S.; Moon, W.; Wilson, W. L.; West, J. E.; Yu, S. M., Permanent polarity and piezoelectricity of electrospun alpha-helical poly(alpha-amino acid) fibers. *Adv Mater* **2011,** *23* (34), 3954-8.
57.	Yucel, T.; Cebe, P.; Kaplan, D. L., Structural Origins of Silk Piezoelectricity. *Adv Funct Mater* **2011,** *21* (4), 779-785.

58. Kitagawa, K.; Morita, T.; Kimura, S., A helical molecule that exhibits two lengths in response to an applied potential. *Angew Chem Int Ed Engl* **2005,** *44* (39), 6330-3.
59. Zhao, M. H.; Wang, Z. L.; Mao, S. X., Piezoelectric characterization of individual zinc oxide nanobelt probed by piezoresponse force microscope. *Nano Lett* **2004,** *4* (4), 587-590.
60. Furukawa, T.; Seo, N., Electrostriction as the Origin of Piezoelectricity in Ferroelectric Polymers. *Jap J Appl Phys Part 1* **1990,** *29* (4), 675-680.
61. Gibaud, T.; Barry, E.; Zakhary, M. J.; Henglin, M.; Ward, A.; Yang, Y.; Berciu, C.; Oldenbourg, R.; Hagan, M. F.; Nicastro, D.; Meyer, R. B.; Dogic, Z., Reconfigurable self-assembly through chiral control of interfacial tension. *Nature* **2012,** *481* (7381), 348-51.

**Table of Contents Graphic:**

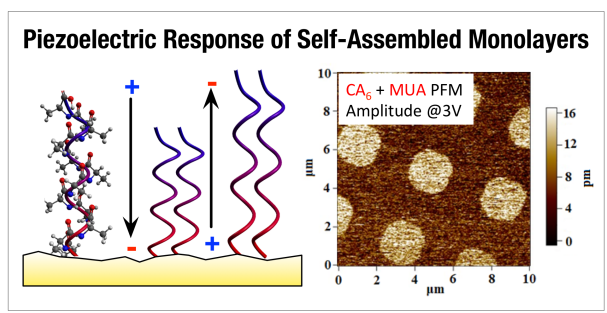